\documentclass[conference,table]{IEEEtran}
\IEEEoverridecommandlockouts
\usepackage{bbding}
\usepackage{cite}
\usepackage{amsmath,amssymb,amsfonts}
\usepackage{bm} 
\usepackage{textcomp}
\usepackage{caption}
\usepackage{graphicx}
\usepackage{subfigure}
\usepackage{pgfplots}
\pgfplotsset{compat=1.18}
\usepackage{subcaption}
\usepackage{multirow}
\usepackage{algorithmic}
\usepackage{algorithm}
\usepackage{hyperref}
\pgfplotsset{ 
cycle list={%
{draw=black,mark=star,solid},
{draw=black, mark=square,solid}}}
\usepackage{breqn}
\usepackage{xcolor}
\usepackage{tikz}
\usetikzlibrary{plotmarks}
\definecolor{darkblue}{rgb}{0.0, 0.0, 0.55}
\definecolor{magenta}{rgb}{0.79, 0.08, 0.48}

\usepackage{caption}
\usepackage{comment}
\usepackage{authblk}
\usepackage{fancyhdr}
\usepackage{subcaption}

\begin{document}
\title{PhD Thesis Summary: Methods for Reliability Assessment and Enhancement of Deep Neural Network Hardware Accelerators}
\author{Mahdi Taheri}

\affil{Tallinn University of Technology, Tallinn, Estonia}
\maketitle
\begin{abstract}

This manuscript summarizes the impact of the work done in the doctoral thesis, introducing novel cost-efficient methods for assessing and enhancing the reliability of DNN hardware accelerators. A comprehensive Systematic Literature Review (SLR) was conducted, categorizing existing reliability assessment techniques, identifying research gaps, and leading to the development of new analytical reliability assessment tools. Additionally, this work explores the interplay between reliability, quantization, and approximation, proposing methodologies that optimize the trade-offs between computational efficiency and fault tolerance. Furthermore, a real-time, zero-overhead reliability enhancement technique, AdAM, was developed, providing fault tolerance comparable to traditional redundancy methods while significantly reducing hardware costs. The impact of this research extends beyond academia, contributing to multiple funded projects, master’s courses, industrial collaborations, and the development of new tools and methodologies for efficient and reliable DNN hardware accelerators.
\end{abstract}

\section{Motivation}\label{chap:introduction}\label{chap:Motivation} 
DNNs are deployed in their target application by different DHA platforms, including Field-Programmable Gate Arrays (FPGAs), Application-Specific Integrated Circuits (ASICs), and Graphics Processing Units (GPUs) \cite{ibrahim2020soft}. Depending on the DHA and the application’s environment, different fault types may threaten the component's reliability \cite{shafique2020robust}. Faults are originated from hardware, however, they can also be modeled at software platforms for ease of study. Accordingly, the reliability of DNNs is tightly coupled with the reliability of DHAs as faults are coming from hardware. It is worth highlighting that the reliability in this thesis does not relate to the reliability in software engineering or security issues e.g., adversarial attacks.

Throughout the literature, various methods of DNN reliability assessment and enhancement are presented. Some review papers have been published on the topic of DNNs reliability enhancement methods \cite{shafique2020robust}. These works aim to formulate the reliability problem in DNNs, categorize available reliability improvement methods in this domain, and overview the fault injection methods for reliability assessment. Reliability assessment of DNNs is a process for evaluating the reliability of a DNN that is being executed either as a software model or by a hardware platform [\ref{pub8}]. The analysis in \cite{torres2017fault} reviews the subject of fault tolerance in DNNs and describes different fault models and reliability improvement methods in DNNs. Subsequent works such as \cite{shafique2020robust} provide extensive reviews on the reliability improvement methods for DNNs and characterize taxonomies of different methods. Nevertheless, they do not consider the assessment and evaluation methods of the reliability of DNNs. Other surveys  \cite{ruospo2023survey} have reviewed fault injection methods for DNNs reliability assessment, with the former work focused merely on fault criticality assessment and the latter included only a few papers in the survey. In [\ref{pub8}], the first Systematic Literature Review (SLR) is presented and dedicated to all methods of reliability assessment of DNNs. 
\subsection{Problem Formulation}\label{chap:Problem Formulation
} 
It has been shown that the functionality of DNNs in terms of accuracy is remarkably degraded in the presence of faults \cite{dos2018analyzing}. 
As there exist no commonly accepted reliability assessment metrics for DNNs, they require more study to make them applicable for safety-critical applications. Existing state-of-the-art fault-tolerant solutions rely on redundant DNNs with implementation diversity. However, this solution is both costly and does not contribute to facilitating reliability assessment for the overall system.

Thus, the main problem this thesis addresses is the reliable operation of DNN hardware accelerators under fault conditions. The aim is to develop novel assessment frameworks that accurately evaluate the impact of faults on DNNs and propose enhancement techniques that mitigate these effects while maintaining computational efficiency. Specifically, this thesis addresses the following problems:

\begin{itemize}

\item \textbf{Cost-Efficient Reliability Assessment} – Developing methods to evaluate the reliability of AI hardware and DNN accelerators with minimal cost and computational overhead.

\item \textbf{Cost-Effective Reliability Enhancement} – Designing efficient techniques to improve the dependability of AI hardware while optimizing for cost and performance trade-offs.

\item \textbf{Accessibility of Reliability Tools for Research community} – Ensuring that reliability assessment and enhancement tools are readily available and usable by the research community to foster advancements in the field.
\end{itemize}

\subsection{Research Objectives}\label{chap:Research Objectives}

Considering the above problems, this PhD thesis aims to address them as follows:

\begin{itemize}
\item Develop and introduce innovative techniques and frameworks for assessing and enhancing the reliability of DNNs.

\item Conduct an in-depth analysis of various reliability and hardware optimization strategies, such as quantization and approximation, to evaluate their collective impact on DNN accuracy, reliability, and hardware performance.

\item Develop and deploy mitigation techniques for DNNs at both the architectural and hardware accelerator levels, including fault-tolerant designs and redundancy methods.
\end{itemize}
\section{Scientific and Technological Excellence}

The contributions of this PhD thesis in tackling the problems explained in subsection 1.A consists of:
\begin{itemize}
\item Providing a comprehensive overview of essential concepts related to DNNs, reliability assessment, and enhancement, including DNN accelerators and optimization techniques for edge AI. [\ref{pub8}]

\item Proposing methodologies for assessing the reliability of DNNs at different abstraction levels, ranging from hardware-accurate, simulation-based fault injection approaches to emulation-based methodologies on actual hardware. 
[\ref{pub1},\ref{pub5},\ref{pub6},\ref{pub},\ref{pub7},\ref{pub9},\ref{pub10},\ref{pub11},\ref{pub}]

\item Introducing various mitigation techniques for DNNs based on the reliability assessment results. These techniques, categorized into DNN architecture and DNN hardware accelerator levels, include fault-tolerant designs and algorithmic-level redundancy methods. [\ref{pub1}, \ref{pub3}, \ref{pub4}, \ref{pub18}, \ref{pub9}, \ref{pub10}]
\end{itemize}

\subsection{Contribution1: Comprehensive survey of reliability assessment and enhancement}
    
This subsection is based on the following paper:
[\ref{pub8}]
\subsubsection{Concept and
approach}
    
In recent years, several studies have been published accordingly to assess the reliability of DNNs. In this regard, various reliability assessment methods have been proposed on a variety of platforms and applications. Hence, there is a need to summarize the state of the art to identify the gaps in the study of the reliability of DNNs. In this work, a Systematic Literature Review (SLR) on the reliability assessment methods of DNNs is conducted to collect relevant research works as much as possible, present a categorization of them, and address the open challenges.

Through this SLR, three kinds of methods for reliability assessment of DNNs are identified including Fault Injection (FI), Analytical, and Hybrid methods. Since the majority of works assess the DNN reliability by FI, this work characterize different approaches and platforms of the FI method comprehensively. Moreover, Analytical and Hybrid methods are propounded. Thus, different reliability assessment methods for DNNs have been elaborated on their conducted DNN platforms and reliability evaluation metrics. Finally, this work highlights the advantages and disadvantages of the identified methods and addresses the open challenges in the research area. it is concluded that Analytical and Hybrid methods are light-weight yet sufficiently accurate and have the potential to be extended in future research and to be utilized in establishing novel DNN reliability assessment frameworks. 
\subsubsection{Obtained results}
    
In this survey, we \textbf{summarized 139 papers} published over the years 2017-2022.

The number of papers based on different reliability assessment methods among all identified works in this literature review were profiled. BAsed on the study, the majority of works use fault injection to assess the reliability of DNNs while only 10\% of the works consider analytical (11 works) and hybrid analytical/FI (3 works) methods. This insight revealed a gap: although analytical approaches offer significant advantages for certain applications, they have been largely overlooked by the research community. To address this, we developed two analytical reliability assessment tools that significantly enhance the speed and accuracy of DNN reliability evaluation [\ref{pub7}], \cite{GENIE}.

\subsubsection{Novelty and foundational characters}
    
This survey represents the first comprehensive Systematic Literature Review (SLR) dedicated to all methods of reliability assessment for Deep Neural Networks (DNNs). Unlike previous works that focus on either fault injection (FI) or reliability enhancement techniques separately, our study systematically categorizes and compares fault injection, analytical, and hybrid reliability assessment methods across different DNN platforms and applications. By reviewing 139 research papers from 2017 to 2022, this paper provide a structured taxonomy that enables researchers and practitioners to identify the most suitable assessment techniques based on their specific application needs. Furthermore, this paper highlight open challenges and gaps in the existing literature, demonstrating that analytical and hybrid methods offer lightweight yet sufficiently accurate alternatives to exhaustive fault injection experiments. This work not only synthesizes state-of-the-art methodologies but also establishes a foundation for developing future reliability assessment frameworks, making it a key reference in the field of DNN hardware reliability research. 
This survey led to more than \textbf{12 scientific papers} that shaped the basis of this PhD thesis and \textbf{four keynote speeches} at renowned international conferences. Our paper is \textbf{downloaded about 5283 times} according to the publisher website and it has received \textbf{34 citations} according to Google Scholar.

\subsection{Contribution2: Interplay of Reliability, Quantization, and Approximation in DNN hardware accelerators}
    
This subsection is based on the following papers: [\ref{pub1}, \ref{pub}, \ref{pub18}].

\subsubsection{Concept and
approach}
    
In practice, deployment of a DNN accelerator for the safety-and mission-critical applications  (e.g., autonomous driving) requires addressing the trade-off between different design parameters of \textit{hardware performance}, e.g., area, power, delay, and \textit{reliability}.

A compromise between conflicting requirements can be achieved by different ways like quantization of network parameters to reduce memory footprint and computing load, which can happen at the DNN model level, and simplifying the implementation at the hardware level to sacrifice the precision of results but benefiting from lower resource utilization, energy consumption, and higher system efficiency. Quantization-aware training, post training quantization and \textit{Approximation Computing (AxC)} are of such concepts \cite{choudhary2022approximate}. Moreover, the assessment of the reliability of DNN accelerators is a challenging issue by itself. 

DNNs are known to be inherently fault-resilient due to the high number of learning process iterations and also several parallel neurons with multiple computation units.
Nevertheless, faults may impact the output accuracy of DNNs drastically \cite{bosio2019reliability}, and in case of resource-constrained critical applications, DNNs' fault resiliency is required to be evaluated and guaranteed. 
The complexity of such evaluation motivates an \textit{automated tool-chain} with Quantization, AxC and resiliency analysis to support \textit{Design Space Exploration (DSE)} for DNN accelerators already at the early design stage, i.e. starting from a high-level description.

To comprehensively study the impact of transient faults, this thesis considered faults in the activations [\ref{pub1}], and faults in the parameters of the network [\ref{pub}, \ref{pub18}] by exploring different quantization and approximation levels in the neural networks.
 \begin{figure}[h]
    \centering
    \includegraphics[width = 0.5\textwidth]{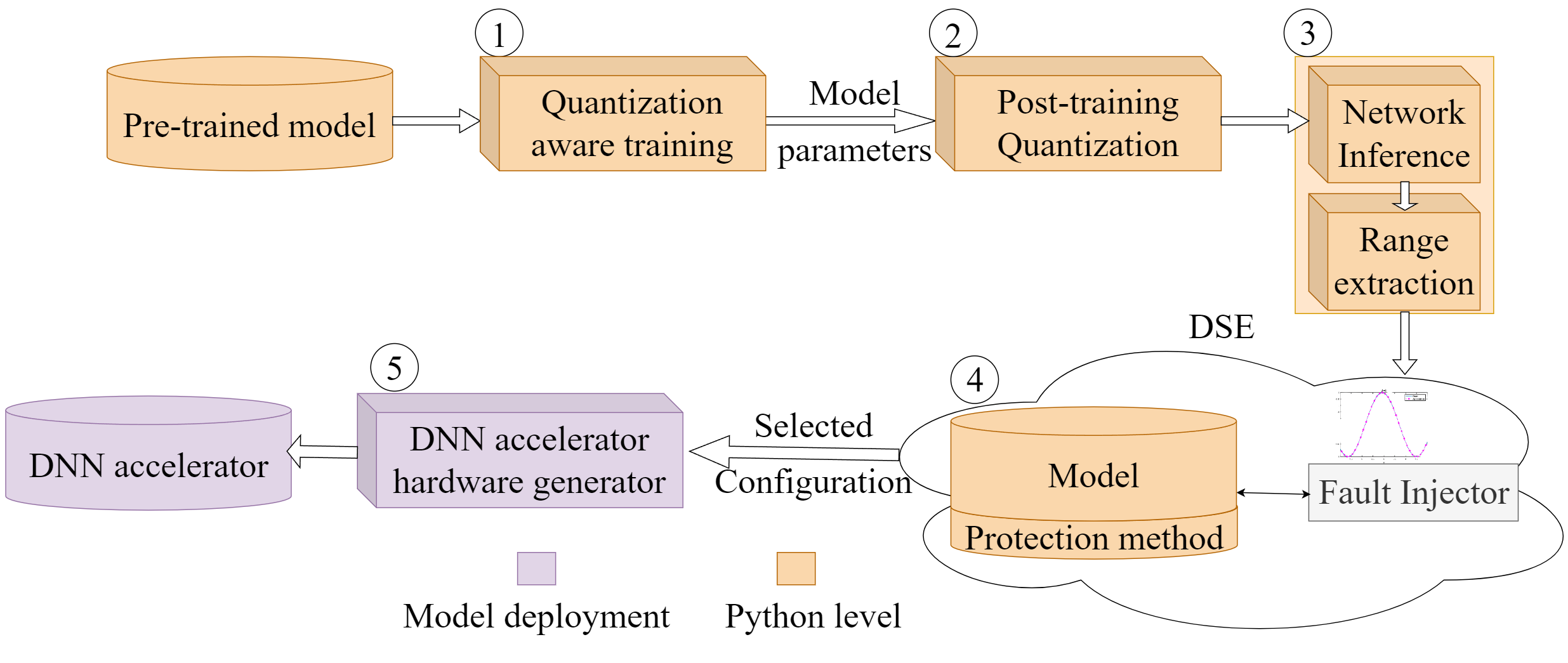}
    \caption{Proposed methodology flow for exploration of activation fault reliability in quantized systolic array based DNN accelerators}
    \label{QRNet_methodology_flow}
\end{figure}

\textbf{Exploration of Activation Fault Reliability in Quantized Systolic Array-Based DNN Accelerators [\ref{pub1}}

This framework uses the high-level description of a DNN as an input and is capable of providing a transient-fault-resilient systolic-array-based FPGA implementation of the network utilizing the design parameters selected by the DSE. The main contributions in this work are as follows:

\begin{itemize}
    \item A methodology for holistic exploration of quantization and reliability trade-offs in systolic-array implementation that enables assessing the trilateral impact of quantization on accuracy, activation fault reliability, and hardware performance. 
    \item A fully-automated framework that is capable of applying quantization-aware training, post-training quantization, range-restriction, fault simulation, and implementing the whole methodology down to hardware implementation to measure actual hardware parameters like area, latency, etc.
    \item A lightweight and effective protection technique is developed and adopted in the framework toolchain to provide the final reliable systolic-array-based FPGA implementation of the network

\end{itemize}

Fig. \ref{QRNet_methodology_flow} illustrates the methodology flow established in the toolchain for reliability and hardware performance analysis of quantized DNN hardware accelerators. 
\textbf{Step 1: Quantization-Aware Training}
The framework applies full quantization to activations, weights, and biases using TFLite, though other libraries can be integrated. The output includes the quantized network’s parameters and architecture.

\textbf{Step 2: Post-Training Quantization}
Further quantization is applied with minimal accuracy loss, supporting up to 4-bit INT quantization. The framework provides the network’s accuracy as a baseline for later steps. The framework focuses on uniform unsigned symmetric quantization.

\textbf{Step 3: Inference and Range Extraction}
Inference is run to extract activation ranges for reliability analysis. These are validated using test data.

\textbf{Step 4: Design Space Exploration}

\textbf{Step 4A: Fault Simulation}
A Python-based systolic-array simulation injects single-bit faults in activations to assess reliability. Multiple-bit faults can be evaluated if needed. The framework measures accuracy degradation and Silent Data Corruption (SDC) rates. Once a quantization scheme is chosen, a fault injector evaluates the Quantized DNN (QDNN) before hardware generation.

\textbf{Step 4B: Fault Mitigation}
Out-of-range outputs are clamped within predefined layer-wise bounds using specialized hardware units \ref{mit}. The protection mechanism compares Multiply-Accumulate (MAC) outputs with stored thresholds and selects a correction:
\begin{figure}[h]
    \centering
    \includegraphics[width=0.25\paperwidth]{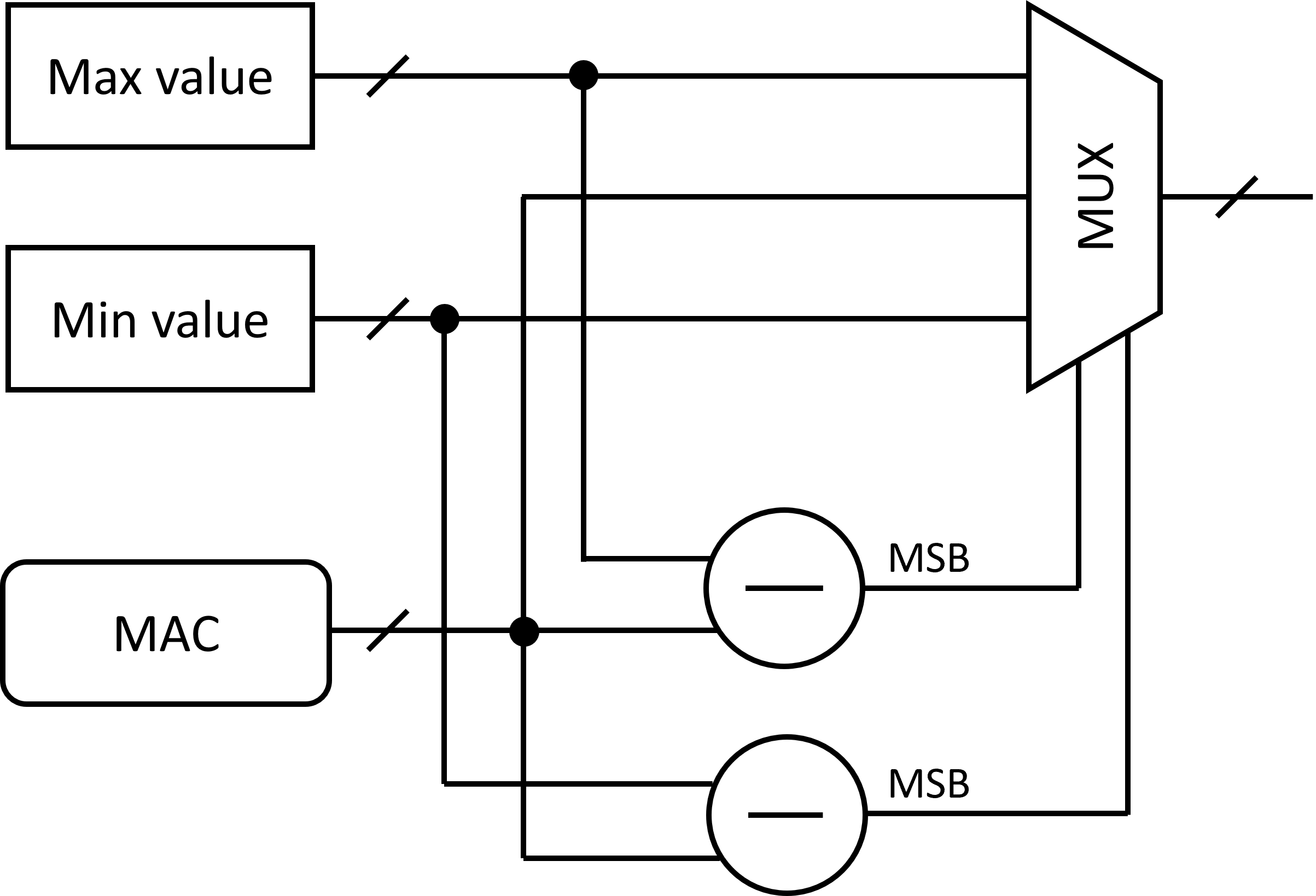}
    \caption{Proposed lightweight mitigation technique}
    \label{mit}
\end{figure}

\textbf{Method 1:} Replace with lower bound.

\textbf{Method 2:} Replace with upper bound.

\textbf{Method 3:} Replace with bound based on MAC output sign.

The approach can be swapped with other protection methods (e.g., FT-ClipAct).

\textbf{Step 5: Hardware Generation}
A systolic-array-based QDNN accelerator is generated for FPGA SoC, optimizing hardware resources. The process includes:

\begin{enumerate}
    \item Determining array size, bit precision, and AXI bandwidth.
    \item Configuring the board with PYNQ for Python and Jupyter Notebook support.
    \item Loading network weights as NumPy arrays and interfacing via Python.
    \item Deploying and running inference on FPGA.
\end{enumerate}

\textbf{FORTUNE: A Negative Memory Overhead Hardware-Agnostic Fault TOleRance TechniqUe in DNNs [\ref{pub18}]}

To address the extensive memory overhead and performance degradation challenges, we present FORTUNE, a model-level hardware agnostic methodology that explores different quantization levels of a DNN model, focusing on reliability, accuracy, memory overhead, and performance aspects. The proposed methodology introduces a novel fault tolerance technique that uses memory savings from quantization to the Most Significant Bit (MSB) within the same memory elements, maintaining QNN reliability even in the presence of memory faults. A comprehensive GPU-based framework implements this methodology.
Specifically, the proposed fault tolerance methodology integrates into the framework’s iterative optimization loop, allowing users to define thresholds that balance model accuracy, reliability, and performance. The extensive experiments show that, without this protection, QNNs suffer significant accuracy degradation in fault-prone environments. The results underscore the importance of the proposed technique to minimize memory utilization and optimize protected QNN execution time on DNN accelerators. In this work, without loss of generality, we reported the results for GPU as an example.

The key contributions of this paper are:

\begin{itemize}
\item \textbf{Negative Overhead Fault Tolerance Technique}: Proposed a protection technique leveraging quantization to triplicate MSB, ensuring robustness against faults.

\item \textbf{Design Space Exploration Framework}: Developed an open-source framework to assess the impact of quantization on QNN reliability, accuracy, memory utilization, and executing time for design space exploration. \url{https://github.com/nilay1400/DNN-Quantization}

\item \textbf{Introduction of \bm{$P_{drop}$} and Reliability-Aware Performance (RAP) metrics}:
 Introduced $P_{drop}$, the probability of accuracy drop over a device's lifetime with various BERs, and RAP, a metric for evaluating trade-offs in fault-prone and resource-constrained environments.

\end{itemize}

In follows, the methodology for quantizing a DNN model is explained, including the reliability  evaluation in fault-prone environments, and the proposed protection technique.

The goal is to achieve a quantized model that not only meets predefined accuracy and reliability thresholds but also provides insights into memory utilization and execution time. The algorithm uses a trained FP32 DNN model as input, along with three key parameters: an accuracy threshold $a$, a reliability threshold $b$, and a quantization range $[m, n]$. The accuracy threshold $a$ represents the minimum acceptable accuracy of the model after quantization in the fault-free model, while the reliability threshold $b$ specifies the maximum permissible drop in accuracy due to potential faults. The quantization range $[m, n]$ defines the range of bit widths to be explored during the quantization process.

\textbf{Reliability Evaluation and Enhancement}. The algorithm proceeds to evaluate its reliability under fault conditions.  Faults are incrementally injected into the quantized model's weights, simulating different Bit Error Rates (BERs). After each fault injection, the model's accuracy is re-evaluated, and the accuracy drop is calculated as the difference between the golden accuracy and the current accuracy.  

In the proposed approach, only the MSB bit is protected by replicating it in two redundant bits. During inference, these redundant bits, along with the protected bit, undergo a majority voting process to determine the final value of the protected bit. This method ensures that even with potential faults, the most critical bit remains reliable.

\begin{figure}
    \centering
    \includegraphics[width=0.35\textwidth]{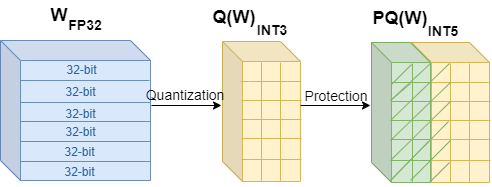}
    \caption{An example of protected 3-bit weights}
    \label{fig:protection}
\end{figure}

Figure\ref{fig:protection} illustrates an example of 3-bit weights with a protected bit. In this example, the FP32 weights are first converted into unsigned 3-bit integer values (shown in orange), and the MSB is replicated into two redundant bits (shown in green). Similarly, in 5-bit weights with one protected bit, two redundant bits are added.  More generally, in $i$-bit quantization with one protection bits, two redundant bits are added, and one comparison is performed during inference. 

\textbf{\bm{$P_{drop}$} and \bm{$RAP$}}\\
Accuracy drop is evaluated independently of any physical effects that faults might have on memory. To account for these effects on accuracy drop, we define 
$P_{drop}$ as the probability of experiencing the accuracy drop during the device's lifetime:
\begin{equation}
    P_{drop} = N^2 \times W^2 \times T / t \times P_{single} \times BER \times acc\_drop
\end{equation}
where $N$ is the number of parameters, $W$ is their bit width, $T$ is device life time, $t$ is test time interval, $P_{single}$ is the probability of one bit flip during $t$ and $acc\_drop$ is the reported accuracy drop in the BER. This metric is based on the probability of a one-bit flip stated in \cite{yan2020single}. As the definition suggests, the more resilient the networks are, the smaller the value of $P_{drop}$ becomes. 

To account for performance along with accuracy drop and memory footprint, we define the Reliability-Aware Performance (RAP) metric as:

\begin{equation}
    RAP = acc\_drop \times mem\_ovh \times perf\_ovh
\end{equation}

where $acc\_drop$ represents the accuracy drop, $perf\_ovh$ refers to the execution time overhead, and $mem\_ovh$ denotes the memory utilization overhead.   Smaller RAP values indicate more reliable networks with lower memory and performance overhead.

\textbf{DeepAxe: A Framework for Exploration of Approximation and Reliability Trade-offs in DNN Accelerators [\ref{pub}]}

High-Level Synthesis (HLS) tools bridge high-level programming and hardware implementation and allow overcoming the complexity of the process and reducing the design time. Recently, DNN-tailored HLS tools were proposed, e.g., CNN2gate\cite{2}. Such tools are capable of providing a synthesizable C implementation of DNNs for FPGAs from a high-level description in a language such as e.g., Keras. 
 \begin{figure}[h]
    \centering
    \includegraphics[width = 0.47\textwidth]{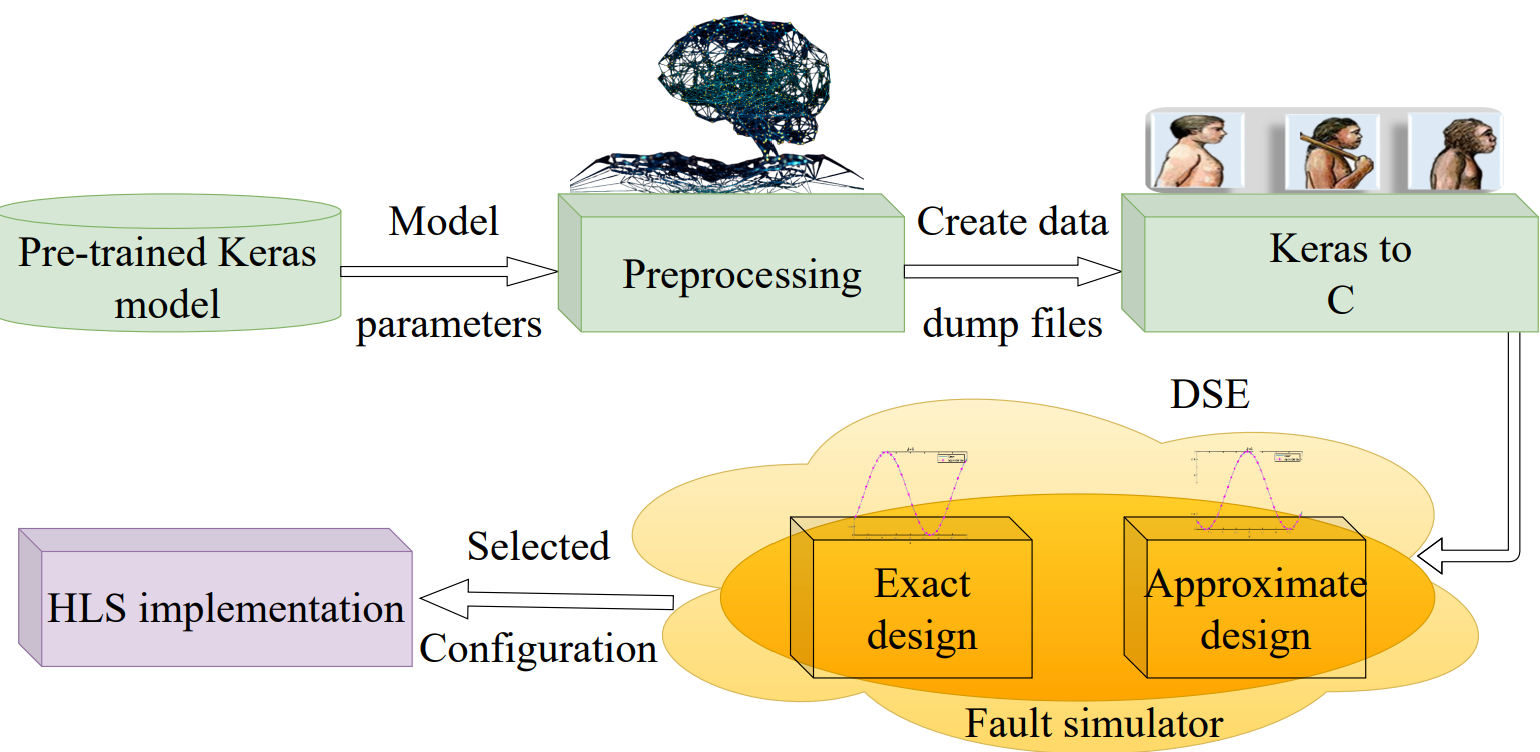}
    \caption{DeepAxe methodology flow}
    \label{method}
\end{figure}
This paper presents a novel framework and a fully automated tool-chain, DeepAxe to provide a design space exploration for FPGA-based implementation of DNN accelerators by analyzing
approximation and soft-error reliability trade-offs. 
To the best of our knowledge, this is the first framework that holistically considers both the transient fault resiliency and hardware performance of DNN accelerators as design parameters. DeepAxe is empowered by techniques for quantizing the networks and providing the capability of substituting the exact computing (ExC) units of the network with AxC units and identifying the optimal design points for selective approximation. 

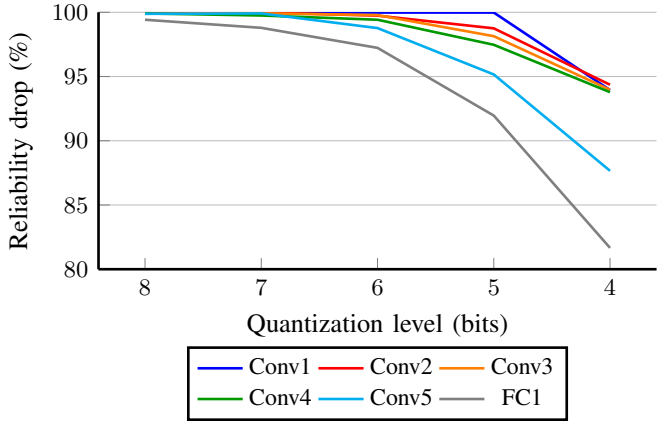
\begin{figure}[!h]
\begin{center}
\begin{tikzpicture}
\begin{axis} [height=5cm, width=9cm, ymin=80, ymax=100, axis lines*=left,
              ymajorgrids=true, line width=1pt,
              x dir=reverse, xtick={8,7,6,5,4},
              enlarge x limits=0.1, tick label style={font=\small},
              xlabel=Quantization level (bits), ylabel={Reliability drop (\%)}, ylabel near ticks,
              legend style={font=\small,at={(0.5,-0.3)},anchor=north,legend columns=3}]
\addplot[color=blue] coordinates {
	(8,99.92)
	(7,99.99)
	(6,99.98)
        (5,99.98)
	(4,93.95)
};
\addplot[color=red] coordinates {
	(8,99.99)
	(7,99.99)
	(6,99.74)
        (5,98.75)
	(4,94.36)
};
\addplot[color=orange] coordinates {
	(8,99.9)
	(7,99.92)
	(6,99.78)
        (5,98.14)
	(4,93.97)
};
\addplot[color=green!60!black] coordinates {
	(8,99.91)
	(7,99.75)
	(6,99.42)
        (5,97.47)
	(4,93.78)
};
\addplot[color=cyan] coordinates {
	(8,99.88)
	(7,99.91)
	(6,98.77)
        (5,95.16)
	(4,87.66)
};
\addplot[color=gray] coordinates {
	(8,99.42)
	(7,98.79)
	(6,97.23)
        (5,91.96)
	(4,81.66)
};
\legend{Conv1,Conv2,Conv3,Conv4,Conv5,FC1}
\end{axis}
\end{tikzpicture}
\end{center}
\caption{AlexNet layer-level reports of reliability drop (\%) based on different quantization levels (unprotected design)}
\label{re_improv0}
\end{figure}

\begin{figure*}[h!]
\centering
{\includegraphics[width=0.45\textwidth]{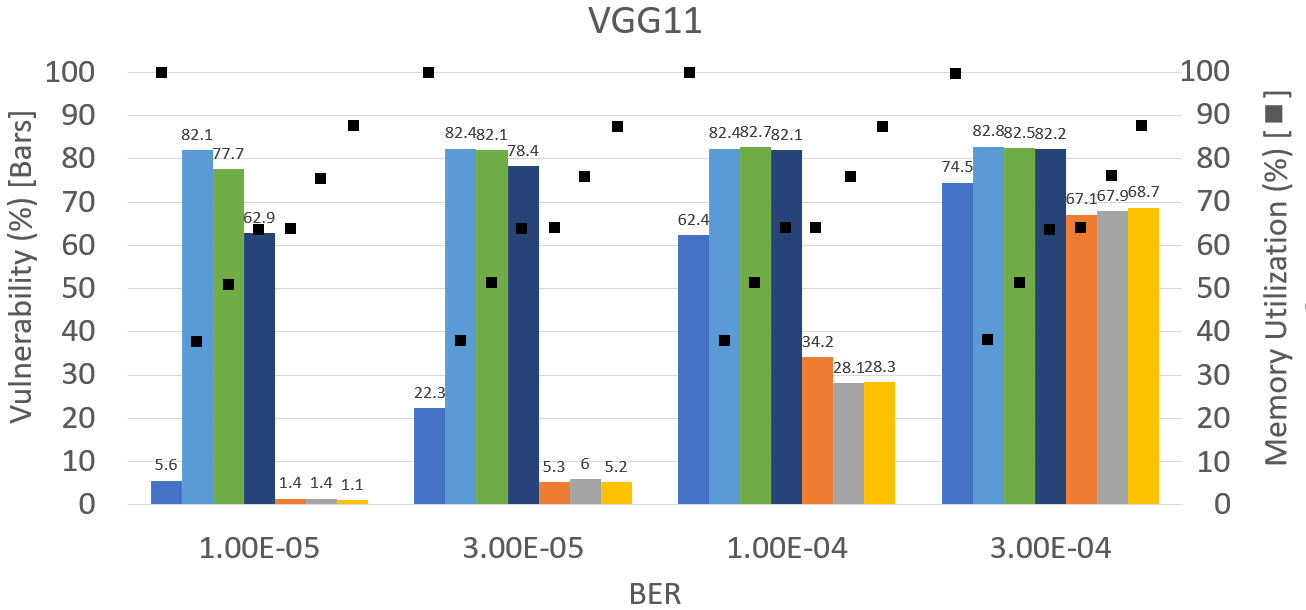}}%
\hfill
{\includegraphics[width=0.45\textwidth]{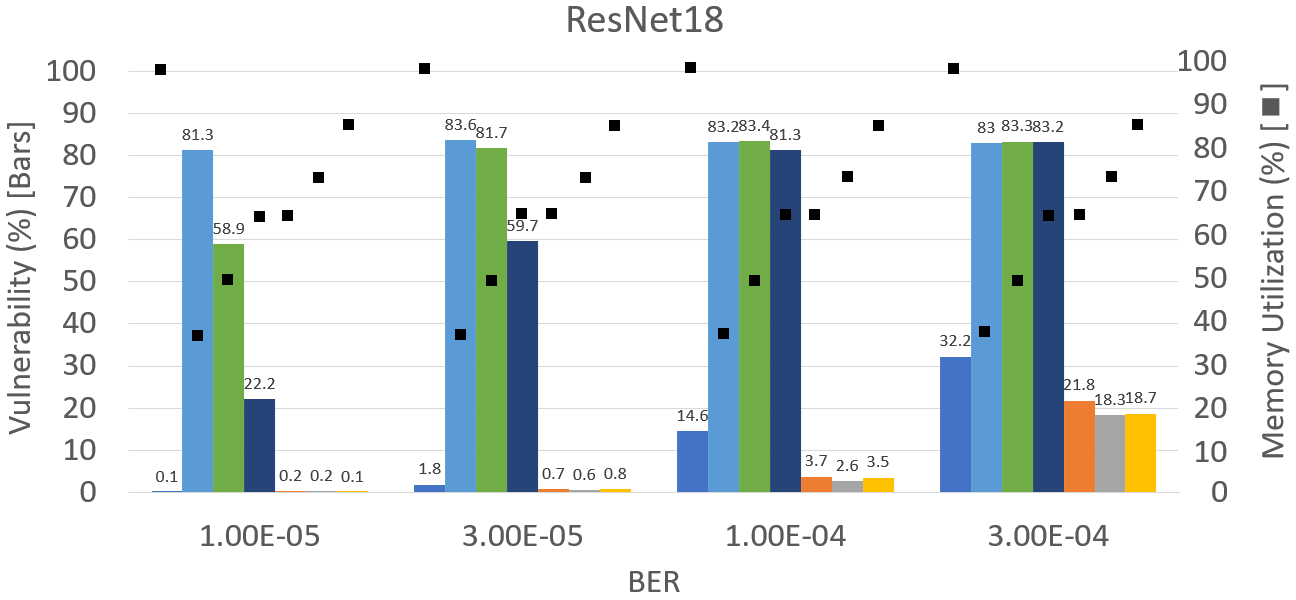}}
\hfill
{\includegraphics[width=0.45\textwidth]{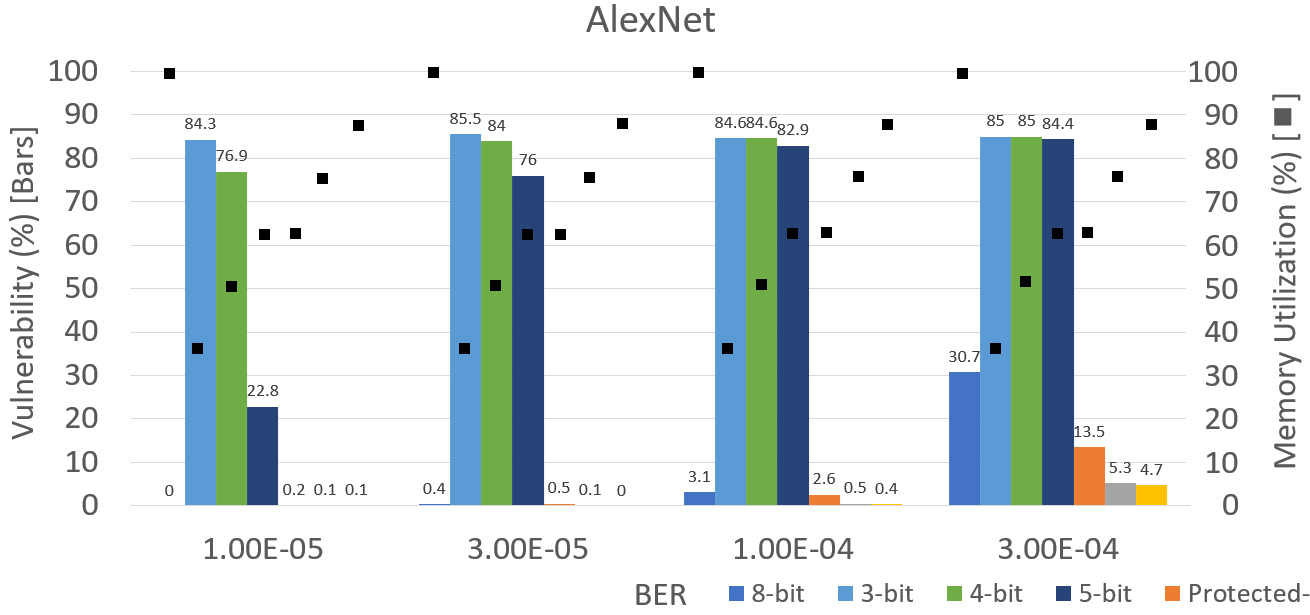}}%
\hfill
{\includegraphics[width=0.45\textwidth]{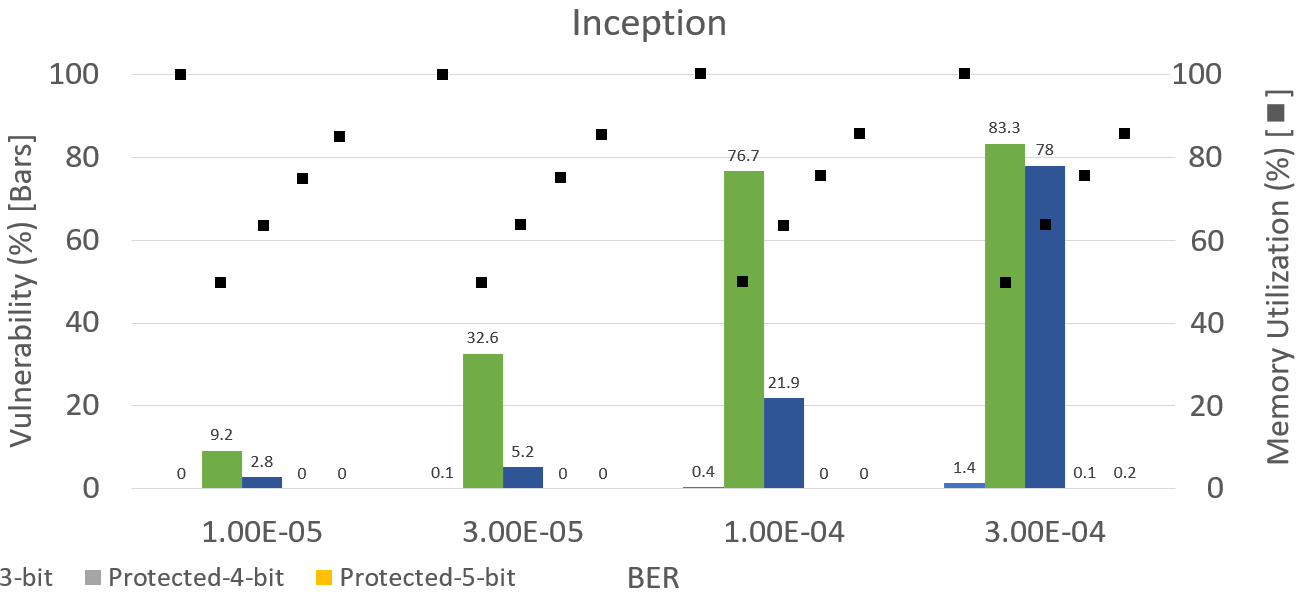}}
\hfill
\caption{Vulnerability (accuracy drop due to fault injection) and memory utilization trade-offs in different benchmarks: VGG-11 (CIFAR), ResNet-18 (CIFAR), AlexNet (FashionMNIST) and Inception(CIFAR). }
\label{fig:trade-off}
\end{figure*}
\begin{table*}
    \centering
    \caption{Memory Utilization, Execution Time, Vulnerability (accuracy drop due to fault injection) and $P_{drop}$ in DNNs.}
    \begin{tabular}{|c|c|c|c|c|c|c|c|c|c|c|c|} \hline 
       \multirow{2}{*}{Model}  & \multirow{2}{*}{Type} & \multirow{2}{*}{\begin{tabular}[c]{@{}c@{}}Memory\\ Utilization\end{tabular}} & \multirow{2}{*}{\begin{tabular}[c]{@{}c@{}}Executuin\\ Time (\%)\end{tabular}} &  \multicolumn{4}{|c|}{\begin{tabular}[c]{@{}c@{}}Vulnerability (\%)\\ \{BER\} \end{tabular}} & \multicolumn{4}{|c|}{\begin{tabular}[c]{@{}c@{}}$P_{drop}$\\ \{BER\} \end{tabular}}\\ \cline{5-12}
        & & & & 1.00E-5& 3.00E-5&1.00E-4 &3.00E-4 & 1.00E-5& 3.00E-5&1.00E-4 &3.00E-4\\ \hline
 \multirow{4}{*}{VGG11} & 8-bit&225,064,448 & 100& 5.62& 22.30& 62.40&74.52 &50.21E-3 &150.63E-3 &502.10E-3 &1506.30E-3\\ \cline{2-12}
 & P-5-bit& 196,931,392&141.72 &1.11 & 5.23& 28.33&68.73 &35.45E-3 & 106.35E-3& 354.50E-3&1063.52E-3\\ \cline{2-12}
  & P-4-bit&168,798,336 & 141.72&1.45 & 6.01& 28.19&67.99 &25.76E-3 &77.30E-3 & 257.66E-3&773.00E-3\\ \cline{2-12}
  & P-3-bit& 140,665,280&141.72 &1.49 &5.35 &34.20 &67.13 &17.66E-3 &53.00E-3 & 176.68E-3&530.05E-3\\ \hline 
 \multirow{4}{*}{ResNet18} & 8-bit& 66,991,616&100 & 0.18& 1.85& 14.64&32.29 &1.92E-3 & 5.78E-3&19.27E-3 &57.82E-3\\ \cline{2-12} 
 &P-5-bit &58,617,664 & 119.54& 0.16& 0.81&3.54 &18.77 &0.85E-3 &2.57E-3 & 8.58E-3&25.74E-3\\ \cline{2-12}
 & P-4-bit & 50,243,712 & 119.54 & 0.24 & 0.66 & 2.63 & 18.34 &0.61E-3 &1.84E-3 &6.15E-3 &18.47E-3\\ \cline{2-12}
  &P-3-bit &41,869,760 & 119.54& 0.25&0.79 &3.74 &21.89 & 0.51E-3&1.53E-3 & 5.10E-3&15.31E-3\\  \hline 
 \multirow{4}{*}{AlexNet} & 8-bit & 466,316,032 & 100 & 0.00 & 0.49 & 3.14 &30.71 & 88.82E-3&266.47E-3 &888.23E-3 &2664.70E-3\\ \cline{2-12}
 & P-5-bit & 408,026,528 & 102.94 & 0.14 & 0.04 & 0.48 & 4.73&10.47E-3 &31.43E-3 &104.78E-3 &314.46E-3\\ \cline{2-12}
 & P-4-bit & 349,737,024 & 102.94 & 0.19 & 0.10 & 0.52 & 5.30&8.62E-3 &25.87E-3 & 86.24E-3&258.72E-3\\ \cline{2-12}
 & P-3-bit & 291,447,520 & 102.94 & 0.27 &  0.58& 2.62 &13.58 &15.34E-3 & 46.03E-3& 153.44E-3&460.33E-3\\ \hline 
 \multirow{3}{*}{Inception} & 8-bit & 173,011,456 & 100 & 0.09 & 0.16 & 0.45 & 1.43& 0.57E-3& 1.71E-3& 5.71E-3&17.15E-3\\ \cline{2-12}
& P-5-bit & 151,234,496 & 291.73 & 0.00 & 0.02 & 0.02 & 0.19&0.06E-3 & 0.18E-3& 0.60E-3&1.80E-3\\ \cline{2-12} 
& P-4-bit & 129,758,592 & 291.73 & 0.01 & 0.05 & 0.08 & 0.14&0.03E-3 & 0.09E-3& 0.33E-3&0.99E-3\\ \hline 
    \end{tabular}
    \label{tab:all_data}
\end{table*}
DeepAxe uses the Keras description of a DNN as the input and is capable of providing an FPGA-ready approximated and transient-fault-resilient inference implementation of the network based on the design parameters selected based on the DSE results. The main contributions in this work are as follows:
\begin{itemize}
    \item A methodology for selective approximation of reliability-critical DNNs providing a set of Pareto-optimal DNN implementation design space points for the target resource utilization requirements. 
    \item A framework DeepAxe for holistic exploration of approximation and reliability trade-offs in DNN accelerator FPGA-based implementation that enables assessing the trilateral impact of approximation on accuracy, reliability, and hardware performance.
    \item Integration of the fully automated DeepAxe tool-chain into the DeepHLS environment.
\end{itemize}
\begin{figure*}[h!]
    \centering
{\includegraphics[width=0.6\textwidth]{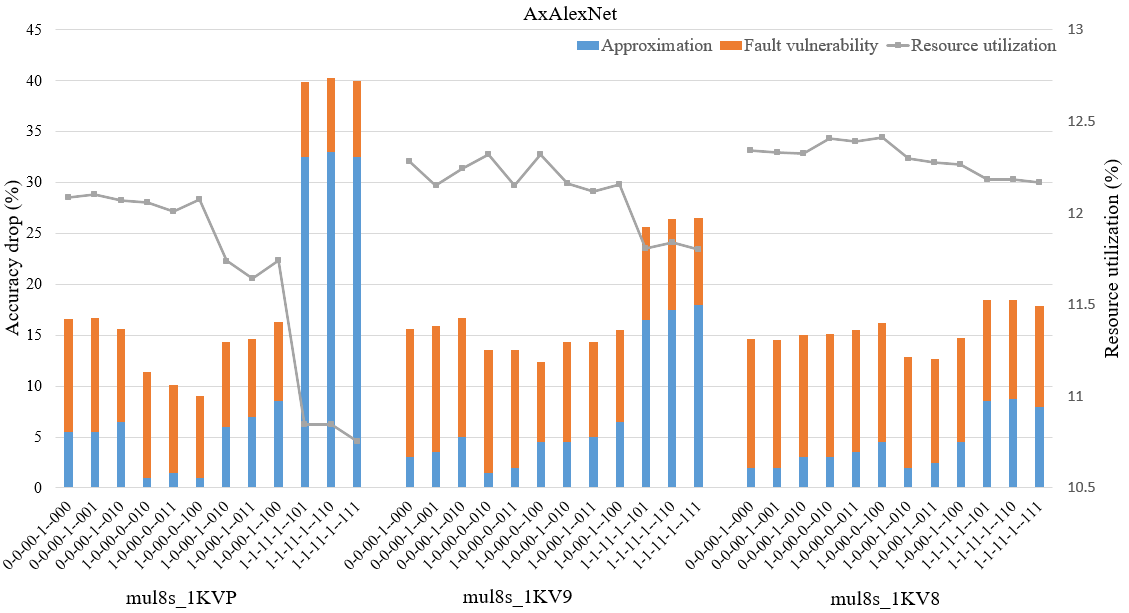}}     
    \caption{Reports of accuracy drop (due to approximation for different configurations), fault vulnerability, and resource utilization of AlexNet}
    \label{ax-accuracy}
\end{figure*}

Fig. \ref{method} illustrates the methodology flow established in the DeepAxe tool-chain for reliability and hardware performance analysis of approximated DNN hardware accelerators. DeepAxe is a framework taking the DNNs' \emph{Pre-trained Keras model} description as the input. Then, DeepAxe feeds the extracted model parameters through the flow to apply the initialization needed before creating the C code. The design, training and test of the DNNs are performed in Python, the \emph{Preprocessing} step is seamlessly integrated into the same environment and is responsible for extracting the required data for the next step.
\subsubsection{Obtained results}
In this section, the results are presented for all the three mentioned works seperately.

\textbf{Experimental results for part 1.1:} Fig. \ref{re_improv0} reports the reliability drop without the protection techniques to show the impact of faults in activations, on different quantization level and layers of an AlexNet network. 

The results shows reliability improvement of the AlexNet network in the presence of a fault of more than 51.79\% in the worst case. Improvements in fault criticality for both networks at the model level are also reported, which demonstrates the positive impact of the protection technique on reducing the criticality of faults in both networks. These data also showcase the increasing fault criticality in different networks by increasing the level of quantization. Based on the results, protection Method~3, which shows the best results for improving reliability among all of the proposed protection techniques, introduces less than 10\% overhead compared to the LUTs required for the unprotected network implementation. Meanwhile, full protection of the network with TMR (Triple Module Redundancy) introduces more than 200\% hardware overhead.

\textbf{Experimental results for part 1.2:}
To evaluate reliability, a random fault injection is conducted across all weights in the DNNs under study. The number of injected faults is determined using a BER ranging from $10^{-5}$ to $3\times10^{-4}$, covering a comprehensive range of potential errors. 
For each bit width and BER, the resulting drop in accuracy (with respect to the corresponding fault-free model) is reported in Table \ref{tab:all_data} as Vulnerability of the networks. Although the protected versions of AlexNet and ResNet-18 do not exhibit significant differences in Vulnerability values at lower BERs, a considerable difference becomes apparent at higher BERs. Conversely, VGG11 and Inception experience less vulnerability across all protected versions and BER levels. Additionally, the unprotected versions of the quantized networks show worst vulnerabilities. $P_{drop}$ values for the DNNs under study are reported in Table \ref{tab:all_data}. Vulnerability and memory utilization tradeoffs are demonstrated in Fig. \ref{fig:trade-off}. The results show the significant improvement of the reliability in the protected networks while keeping the memory utilization in the same range.

\textbf{Experimental results for part 1.3:} Fig. \ref{ax-accuracy} depicts the impact of different approximation units on the case-study DNNs' accuracy, resource utilization and fault vulnerability. For each network, three approximation units are chosen. For approximating the networks, the same configurations are picked to observe the impact of different AxM on the networks. Then all approximation units are applied, and the accuracy drop, fault vulnerability and resource utilization are reported. The correlation between the AxM error metrics, their area overhead, and the accuracy drop of the AxDNN impacted by AxMs lead us toward a conclusion that the network accuracy is generally impacted by a) the level of approximation and the configuration of the layers that are substituted by AxM; b) the error metrics of the AxM that is used as a substitution of ExC unit.

\subsubsection{Novelty and foundational characters}
    
In our proposed work, we derived new insights into the impact of approximation and quantization on the reliability, efficiency, and memory utilization of DNN accelerators. We introduced novel techniques to safeguard the memory and activations of neural networks against transient faults and demonstrated their effectiveness through our results. Additionally, all proposed methods are integrated into automated toolchains and open-source frameworks, enhancing accessibility for the research community and facilitating reproducibility and further studies.

\subsection{Contribution3: Real-time zero overhead reliability enhancement of dnn accelerators}
    
This subsection is based on the following papers:
[\ref{pub3}, \ref{pub4}]

\subsubsection{Concept and
approach}
    
Deep Neural Networks (DNNs) demonstrated a significant improvement in accuracy by adopting computationally intense models. Consequently, the size of these models has increased drastically, imposing challenges in their deployment on resource-constrained platforms \cite{Babak}.
\begin{figure}[h]
    \centering
    \includegraphics[width = 0.25\textwidth]{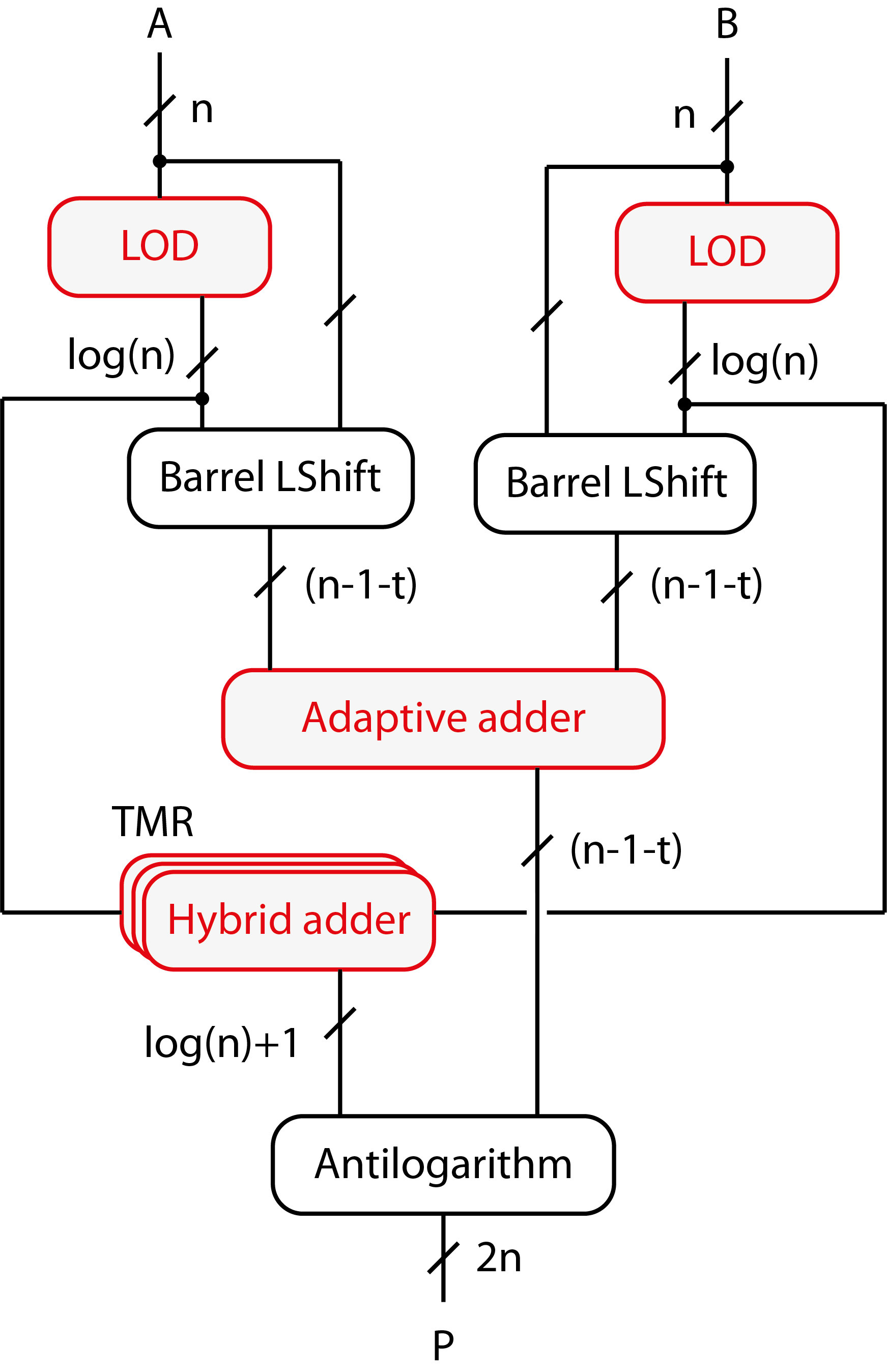}
    \caption{AdAM architecture (the contributions and extensions to the logarithmic Mitchell multiplier are marked with red color)}
    \label{fig:mult}
\end{figure}
\begin{figure}[ht!]
\centering
{\includegraphics[width=0.5\textwidth]{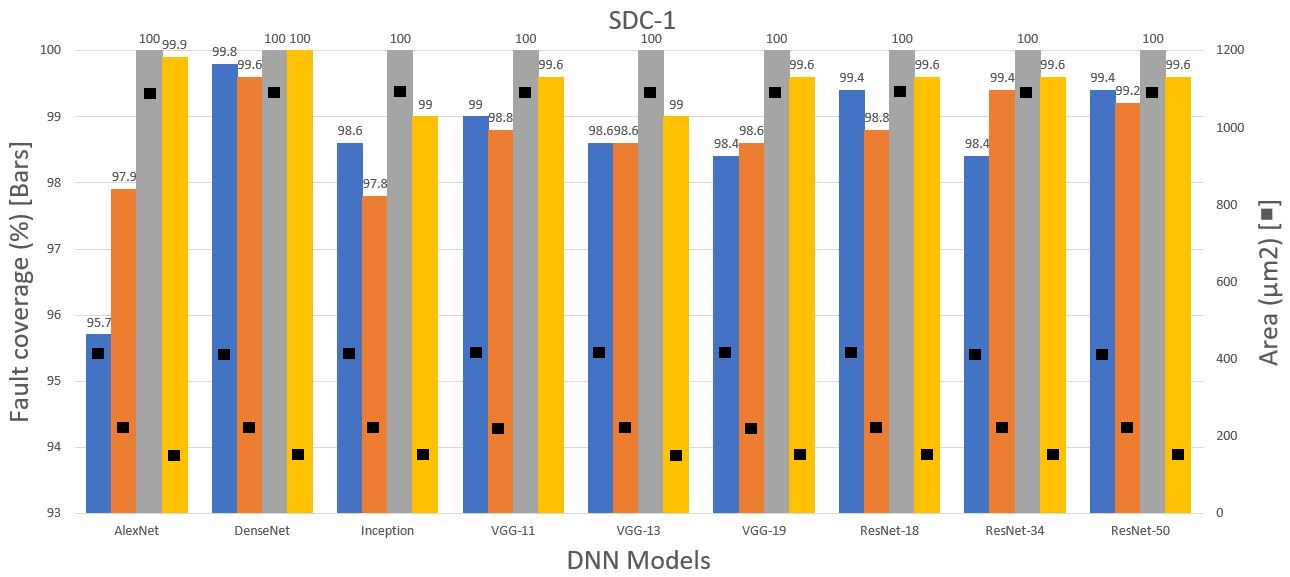}}%
\hfill
{\includegraphics[width=0.5\textwidth]{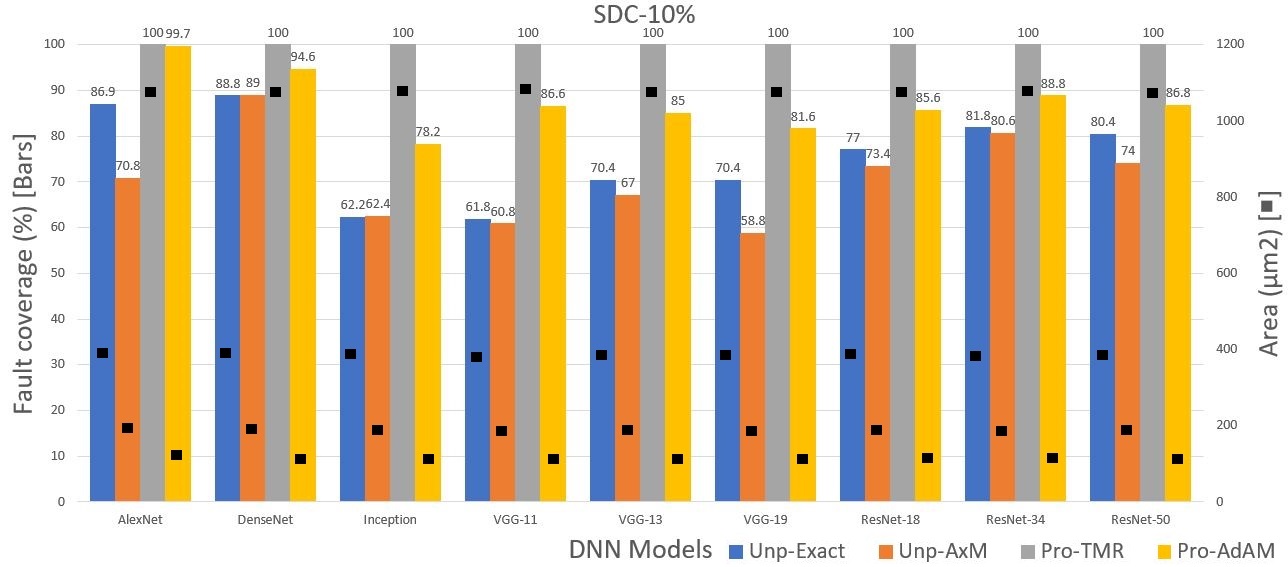}}%
\caption{Hardware efficiency (area) and fault resilience (fault coverage considering SDC-1 and SDC-10\%) trade-offs in different benchmarks: AlexNet (MNIST), DenseNet (CIFAR), Inception (CIFAR), ResNet-18 (CIFAR), ResNet-34 (CIFAR), ResNet-50 (CIFAR), VGG-11 (CIFAR), VGG-13 (CIFAR), VGG-16 (CIFAR). Unp-Exact: unprotected exact multiplier, Unp-AxM: unprotected approximate multiplier, Pro-TMR: exact multiplier protected with TMR, Pro-AdAM: proposed multiplier}
\label{re}
\end{figure}
On the other hand, the role of DNNs in a wide range of safety- and mission-critical applications e.g., autonomous driving, is expanding. Therefore, deploying a DNN accelerator requires addressing the trade-off between different design parameters and \textit{reliability} \cite{taheri2023deepaxe}. Although DNNs possess certain intrinsic fault-tolerant and error-resilient characteristics, it is insufficient to conclude the reliability of DNNs without considering the characteristics of the corresponding hardware accelerator. 

This work presents an architecture of an adaptive fault-tolerant approximate multiplier (AdAM) tailored for ASIC-based DNN accelerators.
The multiplier is based on the logarithmic Mitchell’s multiplier that substitutes multiplication with the addition of approximated logarithms of the operands. The proposed multiplier protects higher-order bits of the product based on the maximum position of the leading one bit in the input strands of the multiplier. This multiplier is a negative overhead fault tolerance approximate multiplier compared to the exact multipliers. The contributions of the paper are as follows:
\begin{itemize}
    \item The architecture of a novel adaptive fault-tolerant approximate multiplier tailored for DNN accelerators, including an adaptive adder relying on an unconventional use of the leading one position value of the inputs for fault detection through optimizing unutilized adder resources.
    
    \item Implementation and validation of the multiplier design.
    
    \item Reliability and hardware performance trade-off assessment and comparison of the proposed multiplier with exact and approximate state-of-the-art multipliers using several state-of-the-art DNN benchmarks.
\end{itemize}

The proposed multiplier provides a reliability level close to the multipliers protected by Triple Modular Redundancy (TMR) while utilizing 2.74$\times$ less area and having 39\% less power-delay product compared to the unprotected exact multiplier. In fact, it has similar area, delay, and power consumption parameters compared to the state-of-the-art approximate multipliers with similar accuracy while providing fault detection and mitigation capability.

\begin{table}[h]
\caption{Accuracy and efficiency of 16-bit approximate multipliers compared with the proposed method}
\label{tab:mult16}
\begin{center}
\begin{tabular}{|c|c|c|c|c|c|c|}
\hline
\begin{tabular}[c]{@{}c@{}}Multiplier\\ Architecture\end{tabular} &
  \begin{tabular}[c]{@{}c@{}}Delay\\ (ns)\end{tabular} &
  \begin{tabular}[c]{@{}c@{}}Power\\ (mW)\end{tabular} &
  \begin{tabular}[c]{@{}c@{}}Area\\ ($\mu m^2$)\end{tabular} &
  \begin{tabular}[c]{@{}c@{}}MARE\\ (\%)\end{tabular} & \multicolumn{1}{l|}{FT} &
  \begin{tabular}[c]{@{}c@{}}PDP\\ (pJ)\end{tabular}\\ \hline
\cellcolor{gray!25}Exact (Wallace)      & \cellcolor{gray!25}1.22 & \cellcolor{gray!25}2.08 & \cellcolor{gray!25}1785 & \cellcolor{gray!25}0.00 & \cellcolor{gray!25}No & 2537.6\cellcolor{gray!25}\\ \hline
DRUM(3)      & 0.88 & 0.13 & 257 & 11.9 & No & 114.4\\ \hline
TOSAM(0,2)   & 0.74 & 0.16 & 342 & 10.9 & No & 118.4\\ \hline
ScaleTrim(3,4) & 1.35 & 0.20 & 281 & 9.23 & No & 279.0 \\ \hline
TOSAM(0,3)   & 0.84 & 0.21 & 423 & 7.6  & No & 176.4\\ \hline
DRUM(4)      & 1.12 & 0.27 & 381 & 5.9  & No & 302.4\\ \hline
ScaleTrim(7,0) & 2.38 & 0.36 & 492 & 4.06 & No & 871.3 \\ \hline
TOSAM(1,5)   & 1.00 & 0.35 & 532 & 4.0  & No & 350.0\\ \hline
\cellcolor{green!25}AdAM(6,7)     &  
1.00\cellcolor{green!25}   & 0.10\cellcolor{green!25} & 440\cellcolor{green!25} & \cellcolor{green!25}3.97
& \cellcolor{green!25}Yes & 100.0\cellcolor{green!25}
\\ \hline

\cellcolor{green!25}AdAM(4,7)     &   1.06\cellcolor{green!25}   & 0.13\cellcolor{green!25} & 451\cellcolor{green!25} & \cellcolor{green!25}3.87
& \cellcolor{green!25}Yes & 137.8\cellcolor{green!25}
\\ \hline
\cellcolor{green!25}AdAM(4,4)     &   0.96\cellcolor{green!25}   & 0.12\cellcolor{green!25} & 434\cellcolor{green!25} & \cellcolor{green!25}3.87
& \cellcolor{green!25}Yes & 115.2\cellcolor{green!25}
\\ \hline
\cellcolor{green!25}AdAM(2,7)     &   
1.32\cellcolor{green!25}   & 0.15\cellcolor{green!25} & 495\cellcolor{green!25} & \cellcolor{green!25}3.97
& \cellcolor{green!25}Yes & 171.6\cellcolor{green!25}
\\ \hline
\cellcolor{green!25}AdAM(2,4)     &   1.13\cellcolor{green!25}   & 0.13\cellcolor{green!25} & 451\cellcolor{green!25} & \cellcolor{green!25}3.85
& \cellcolor{green!25}Yes & 146.9\cellcolor{green!25}
\\ \hline
DRUM(5)      & 1.36 & 0.43 & 532 & 2.9  & No & 584.8\\ \hline
ScaleTrim(9,0) & 2.71 & 0.43 & 541 & 2.2 & No & 1170.6 \\ \hline
TOSAM(2,6)   & 1.21 & 0.38 & 564 & 2.1  & No & 459.8\\ \hline
\end{tabular}
\end{center}
\end{table}
The proposed architecture for adaptive fault-tolerant approximate multiplier tailored for DNN accelerators. This architecture includes an adaptive adder relying on an unconventional use of input Leading One Detector (LOD) values for fault \textit{detection} and \textit{mitigation} through the optimization of unutilized adder resources. A gate-level optimized LOD design and a lightweight triplicated hybrid adder design are used to enhance further the proposed architecture's reliability, resource utilization, and efficiency.
The base for the proposed multiplier is the classical Mitchell multiplier \cite{Mitchell}. However, the methodology can be applied to all logarithmic approximate multipliers. Another level of approximation is introduced in the adaptive adder (Fig.~\ref{fig:mult}) considering the application of this multiplier in DNNs with a proven negligible impact on the network accuracy.

\subsubsection{Obtained results}
    
Table \ref{tab:mult16} reports the accuracy, efficiency, and fault tolerance (FT) of 16-bit approximate multipliers compared with the proposed method. Wallace, DRUM \cite{hashemi2015drum}, TOSAM \cite{vahdat2019tosam}, and ScaleTrim \cite{farahmand2023scaletrim} are used for this comparison. Logarithmic AxMs such as TOSAM show better performance compared to other state-of-the-art non-logarithmic AxMs, including DSM \cite{narayanamoorthy2014energy}, ROBA \cite{zendegani2016roba}, and LETAM \cite{vahdat2017letam}. For instance, TOSAM structures exhibit significantly lower energy consumption and area utilization compared to DRUM, DQ4:2C4, and AS-RoBA multipliers, while maintaining comparable accuracy. Specifically, the TOSAM(1,5) configuration achieves 4\% lower delay, 78\% lower area, and 70\% lower energy consumption compared to the RoBA multiplier. The results also show that the overall impact of the TOSAM multiplier, one of the baseline approaches in this paper, on delay, power, area, and MARE, are lower than those of other approximate multipliers with similar MARE values. Therefore, to respect the page limits of the paper, we did not include other approximate multipliers in our comparison.

The proposed multiplier has similar hardware parameters to the state-of-the-art approximate multipliers with similar accuracy while providing reliability improvement with fault detection and mitigation capability.

To quantify the effectiveness of the proposed method in mitigating faults, \textit{fault coverage} metric is used. It is defined as $(100 - \operatorname{SDC\ value})$ representing the percentage of faults that are correctly handled (i.e., do not result in silent data corruption) by each method.

The proposed architecture of the multiplier is presented in Fig. \ref{fig:mult} (the contributions and extensions to the logarithmic Mitchell multiplier are marked with red color). First, a novel optimized \textit{Leading One Detector (LOD)} circuit is used to find the index of the first ‘1’ bit in each operand. This index denoted as $k$, is the characteristic or integer part of the logarithm and has $log_2(n)$ bits. Then, the operands are shifted left by $k$ bits, aligning the leading one with the Most Significant Bit (MSB). $(n – 1)$ bits after the leading one represent the mantissa part denoted as $m$. The mantissa is truncated to $(n - 1 - t)$ bits.

The truncated operands are passed to the adaptive $(n – 1)$-bit adder that adds mantissa together and duplicates the addition of several higher-order bits depending on the $k$ value of the biggest operand for fault \textit{detection} and \textit{mitigation}. The number of duplicated bits depends on truncation parameter $t$ and duplication level $h$.

Fig. \ref{re} demonstrates the fault tolerance comparison and reliability improvement (for SDC-1 and SDC-10\% as two examples) of different networks by using the protected approximate multiplier proposed in this work compared to the unprotected exact and approximated networks, and protected networks using TMR. As illustrated, TMR has 100\% of protection, but it also requires about 200\% of area overhead. Different from TMR, in our technique we introduce a high-reliability improvement without introducing hardware overhead. 
\subsubsection{Novelty and foundational characters}
    
This paper proposes an architecture of a novel adaptive fault-tolerant approximate multiplier tailored for ASIC-based DNN accelerators. AdAM employs an adaptive adder that relies on an unconventional use of input Leading One Detector (LOD) values for fault detection by optimizing unutilized adder resources. A gate-level optimized LOD design is also proposed to improve the hardware performance as part of the adaptive multiplier. The proposed architecture uses a lightweight fault mitigation technique that sets the detected faulty bits to zero. 

It is demonstrated that the proposed architecture enables a multiplication with a reliability level close to the multipliers protected by TMR while at the same time utilizing 2.74$\times$ less area and with 39.06\% less power-delay product compared to the exact multiplier.

Following the core contributions of this thesis, the proposed methodologies were further extended through subsequent works on adaptive fault resilience and automated architecture search for early-exit DNNs, input-difficulty-aware adaptive early-exit mechanisms, globally guided pruning and sparsification strategies, genetic-algorithm-based reliability assessment, reliability-aware heterogeneous quantization, unified reliability and security enhancement of quantized DNNs, and hardware-aware fast approximation frameworks, representing direct continuations and generalizations of the thesis outcomes
[\ref{pub18},\ref{pub19},\ref{pub20},\ref{pub21},\ref{pub22},\ref{pub23},\ref{pub24},\ref{pub25}].

\section{Impact}

The research presented in this thesis has had a substantial impact in both academic and industrial domains, contributing to multiple funded projects, academic initiatives, and future research directions.
\subsection{Strengthening Industrial Competitiveness and Solving Bottlenecks}

The approaches introduced in this thesis are being considered for further application in industrial environments, such as AI chips at IHP. These advancements aim to accelerate the assessment of industrial-grade accelerators while enhancing the reliability and efficiency of their computing units. By improving performance and reducing processing time, the proposed works directly contribute to the competitiveness and sustainability of AI-driven industries.

\subsection{Contributions to Funded Projects}
The methodologies and findings of this work have been instrumental in various national and international research projects focused on enhancing the reliability and efficiency of Deep Neural Network (DNN) hardware accelerators:

\begin{itemize}
    \item \textbf{Estonian Research Council Grant (PUT PRG1467 - "CRASHLESS")}: Focused on reliable AI accelerators \\
    \item \textbf{EU Grant Project 101160182 ("TAICHIP")}: Addressing trustworthiness in AI chip designs. \\

    \item \textbf{Estonian-French PARROT Project ("EnTrustED")}: Exploring dependable AI. \\

    \item \textbf{Individual Grant (€60,000, VEU22026IA5)}: Co-funded by the European Commission and the Estonian Ministry of Education. \\

    \item \textbf{AI-Disco, 16ME1127)}: funded by Federal Ministry of Research, Technology and Space of Germany (BMFTR). \\

    \item \textbf{THRONE (Deutsche Forschungsgemeinschaft (DFG) - Pending)}: Introducing Dynamic Adaptive DNN Accelerator workloads based on the findings in the presented thesis\\

    \item \textbf{NEUROFABRIX (EXIST-Forschungstransfer - Pending)}: A start-up project based on the findings in the presented thesis on Dependable AI accelerators. \\

\end{itemize}

\subsection{Academic Impact}
The research has also influenced academic initiatives, including:

\begin{itemize}
    \item \textbf{Book Chapter Contribution}: A chapter is in press in an edited volume by Springer, titled \textit{"Machine Learning Systems for High Performance and Dependability: The Role of Hardware Design"}. The chapter title:
    “Cost-efficient reliability of CNN accelerators for Edge“, Springer International Publishing, 2026, pp. 1–29
    \item \textbf{New Master’s Course}: The work has led to the creation of a new master's course on \textit{Special Topics in Deep Neural Networks}, which is being taught at BTU Cottbus from 2025.
    \item \textbf{Foundation for Future Masters´and PhD Research}: This thesis serves as a cornerstone for multiple ongoing and future Master´s (currently 23 active students) and PhD dissertations (currently 3 active students) focused on AI hardware reliability and efficiency.
\end{itemize}

\subsection{Long-Term Vision}
By addressing critical reliability challenges in AI hardware, this work contributes to the broader goal of developing fault-tolerant, and efficient AI accelerators. The research lays the foundation for future advancements in dependable AI systems, influencing both academia and industry in the years to come.

\section{Acknowledgements}
\scriptsize 
This thesis was supervised by Prof. Maksim Jenihhin and Prof. Masoud Daneshtalab, and was financially supported in part by the Estonian Research Council grant PUT PRG1467 "CRASHLESS“, EU Grant Project 101160182 “TAICHIP“, by the Deutsche Forschungsgemeinschaft (DFG, German Research Foundation) – Project-ID "458578717", and by the Federal Ministry of Research, Technology and Space of Germany (BMFTR) for supporting Edge-Cloud AI for DIstributed Sensing and COmputing (AI-DISCO) project (Project-ID "16ME1127").

\section*{List of Publications} 
\addcontentsline{toc}{section}{List of Publications}

 \bibliographystyle{ieeetr}
The present Ph.D. thesis is based on the following publications that are referred to in the text by Roman numbers.\\
\vspace{-2mm}

\begin{publications}

\item\label{pub1} {\textbf{M. Taheri}, N. Cherezova, M. S. Ansari, M. Jenihhin, A. Mahani, M. Daneshtalab, and J. Raik, “Exploration of Activation Fault Reliability in Quantized Systolic Array-Based DNN Accelerators,” in \textit{2024 25th International Symposium on Quality Electronic Design (ISQED)}, 2024.}
\item\label{pub3} {\textbf{M. Taheri}, N. Cherezova, S. Nazari, A. Rafiq, A. Azarpeyvand, T. Ghasempouri, M. Daneshtalab, J. Raik, and M. Jenihhin, “AdAM: Adaptive fault-tolerant approximate multiplier for edge DNN accelerators,” in \textit{2024 IEEE European Test Symposium (ETS)}, 2024.}
\item\label{pub4} {\textbf{M. Taheri}, N. Cherezova, S. Nazari, A. Azarpeyvand, T. Ghasempouri, M. Daneshtalab, J. Raik, and M. Jenihhin, “AdAM: Adaptive Approximate Multiplier for Fault Tolerance in DNN Accelerators,” \textit{ in IEEE Transactions on Device and Materials Reliability, doi: 10.1109/TDMR.2024.3523386},}
\item\label{pub5} {\textbf{M. Taheri}, M. Daneshtalab, J. Raik, M. Jenihhin, S. Pappalardo, P. Jimenez, B. Deveautour, and A. Bosio, “Saffira: a framework for assessing the reliability of systolic-array-based dnn accelerators,” in \textit{2024 27th International Symposium on Design \& Diagnostics of Electronic Circuits \& Systems (DDECS)}, pp. 19--24, 2024.}
\item\label{pub6} {\textbf{M. Taheri}, et al., “Appraiser: Dnn fault resilience analysis employing approximation errors,” in \textit{DDECS}, pp. 124--127, 2023.}
\item\label{pub} {\textbf{M. Taheri}, M. Riazati, M. H. Ahmadilivani, M. Jenihhin, M. Daneshtalab, J. Raik, M. Sjödin, and B. Lisper, “Deepaxe: A framework for exploration of approximation and reliability trade-offs in dnn accelerators,” in \textit{2023 24th International Symposium on Quality Electronic Design (ISQED)}, pp. 1--8, 2023.}
\item\label{pub18} {S. Nazari, \textbf{M. Taheri}, A. Azarpeyvand, M. Jenihhin, M. Daneshtalab, et. al. “FORTUNE: A Negative Memory Overhead Hardware-Agnostic \underline{F}ault T\underline{O}le\underline{R}ance \underline{T}echniq\underline{U}e in D\underline{N}Ns,” In \textit{2024 33th IEEE Asian Test Symposium (ATS 2024).}}
\item\label{pub7} {M. H. Ahmadilivani, \textbf{M. Taheri}, J. Raik, M. Daneshtalab, and M. Jenihhin, “DeepVigor: Vulnerability value ranges and factors for DNNs’ reliability assessment,” in \textit{2023 IEEE European Test Symposium (ETS)}, pp. 1--6, 2023.}
\item\label{pub8} {M. H. Ahmadilivani, \textbf{M. Taheri}, J. Raik, M. Daneshtalab, and M. Jenihhin, “A systematic literature review on hardware reliability assessment methods for deep neural networks,” \textit{ACM Computing Surveys}, vol. 56, no. 6, pp. 1--39, 2024.}
\item\label{pub9} {M. H. Ahmadilivani, \textbf{M. Taheri}, J. Raik, M. Daneshtalab, and M. Jenihhin, “Enhancing Fault Resilience of QNNs by Selective Neuron Splitting,” in \textit{2023 IEEE 5th International Conference on Artificial Intelligence Circuits and Systems (AICAS)}, pp. 1--5, 2023.}
\item\label{pub10} {M. H. Ahmadilivani, M. Barbareschi, S. Barone, A. Bosio, M. Daneshtalab, S. Della Torca, G. Gavarini, M. Jenihhin, J. Raik, A. Ruospo, et al., “Special Session: Approximation and Fault Resiliency of DNN Accelerators,” in \textit{2023 IEEE 41st VLSI Test Symposium (VTS)}, pp. 1--10, 2023.}
\item\label{pub11} {M. H. Ahmadilivani, A. Bosio, B. Deveautour, F. F. Dos Santos, J. D. Guerrero-Balaguera, M. Jenihhin, A. Kritikakou, R. L. Sierra, S. Pappalardo, J. Raik, et al., “Special session: Reliability assessment recipes for dnn accelerators,” in \textit{2024 IEEE 42nd VLSI Test Symposium (VTS)}, pp. 1--11, 2024.}
\item\label{pub12} {M. Jenihhin, \textbf{M. Taheri}, N. Cherezova, M. H. Ahmadilivani, H. Selg, A. Jutman, K. Shibin, A. Tsertov, S. Devadze, R. M. Kodamanchili, et al., “Keynote: Cost-Efficient Reliability for Edge-AI Chips,” in \textit{2024 IEEE 25th Latin American Test Symposium (LATS)}, pp. 1--2, 2024.}
\item\label{pub13} {N. Cherezova, S. Pappalardo, \textbf{M. Taheri}, M. H. Ahmadilivani, B. Deveautour, A. Bosio, J. Raik, and M. Jenihhin, “Heterogeneous Approximation of DNN HW Accelerators based on Channels Vulnerability,” in \textit{IEEE International Conference on Very Large Scale Integration (VLSI-SOC)}, 2024.}

\vspace{-0.25cm}
 \subsection*{\textbf{Other related publications during PhD}} 
\vspace{-0.15cm}

\item\label{pub14} {\textbf{M. Taheri}, “DNN Hardware Reliability Assessment and Enhancement,” \textit{27th IEEE European Test Symposium (ETS)}, May 2022.}

\item\label{pub15} {A. Rezaei, \textbf{M. Taheri}, A. Mahani, and S. Magierowski, “LRDB: LSTM Raw data DNA Base-caller based on long-short term models in an active learning environment,” \textit{arXiv preprint arXiv:2303.08915}, 2023.}

\item\label{pub16} {M. Taheri, \textbf{M. Taheri}, and A. Hadjahmadi, “Noise-tolerance gpu-based age estimation using resnet-50,” \textit{arXiv preprint arXiv:2305.00848}, 2023.}

\item\label{pub17} {\textbf{M. Taheri}, S. Sheikhpour, A. Mahani, M. Jenihhin, “A novel Fault-Tolerant Logic Style with Self-Checking Capability,” \textit{2022 IEEE 28th International Symposium on On-Line Testing and Robust System Design (IOLTS)},.}

\vspace{-0.25cm}
 \subsection*{\textbf{Extracted Publications from thesis}} 
\vspace{-0.15cm}

\item\label{pub18} R. M. Kodamanchili, N. Cherezova, \textbf{M. Taheri}, and M. Jenihhin, “Adaptive fault resilience for early-exit DNNs,” \textit{2025 IEEE 34th Asian Test Symposium (ATS)}, 2025.

\item\label{pub19} \textbf{M. Taheri}, P. Patne, N. Cherezova, A. Mahani, C. Herglotz, and M. Jenihhin, “RL-agent-based early-exit DNN architecture search framework,” \textit{2025 IEEE 28th International Symposium on Design and Diagnostics of Electronic Circuits and Systems (DDECS)}, 2025.

\item\label{pub20} S. Nazari, \textbf{M. Taheri}, A. Azarpeyvand, M. Afsharchi, C. Herglotz, and M. Jenihhin, “Genie: Genetic algorithm-based reliability assessment methodology for deep neural networks,” \textit{2025 11th International Conference on Computing and Artificial Intelligence (ICCAI)}, pp. 264--271, 2025.

\item\label{pub21} S. Nazari, \textbf{M. Taheri}, A. Azarpeyvand, M. Afsharchi, C. Herglotz, and M. Jenihhin, “Reliability-aware performance optimization of DNN HW accelerators through heterogeneous quantization,” \textit{2025 IEEE 26th Latin American Test Symposium (LATS)}, pp. 1--6, 2025.

\item\label{pub22} S. Nazari, M. S. Almasi, \textbf{M. Taheri}, A. Azarpeyvand, A. Mokhtari, A. Mahani, and C. Herglotz, “HAWX: A hardware-aware framework for fast and scalable approximation of DNNs,” \textit{2026 Design, Automation and Test in Europe Conference and Exhibition (DATE)}, Verona, Italy, 2026.

\item\label{pub23} P. Patne, \textbf{M. Taheri}, C. Herglotz, M. Jenihhin, M. Krstic, and M. H{\"u}bner, “DART: Input-difficulty-aware adaptive threshold for early-exit DNNs,” \textit{2026 12th International Conference on Computing and Artificial Intelligence (ICCAI)}, Okinawa, Japan, Apr. 2026.

\item\label{pub24} D. Monachan, S. Nazari, \textbf{M. Taheri}, A. Azarpeyvand, M. Krstic, M. H{\"u}bner, and C. Herglotz, “Mix-and-match pruning: Globally guided layer-wise sparsification of DNNs,” \textit{2026 12th International Conference on Computing and Artificial Intelligence (ICCAI)}, Okinawa, Japan, Apr. 2026.

\item\label{pub25} A. Soltan Mohammadi, S. Nazari, A. Azarpeyvand, \textbf{M. Taheri}, M. Krstic, M. H{\"u}bner, C. Herglotz, and T. Ghasempouri, “RESQ: A unified framework for reliability- and security enhancement of quantized deep neural networks,” \textit{2026 IEEE 27th Latin American Test Symposium (LATS)}, Florian{\'o}polis, Brazil, Mar. 2026.

 \end{publications}

\bibliography{ref}

\end{document}